**V.I. Abrosimov, O.I. Davydovska**

*Institute for Nuclear Research, National Academy of Sciences of Ukraine*

# LOW-ENERGY ISOSCALAR DIPOLE RESPONSE OF HEAVY NUCLEI WITHIN KINETIC MODEL

The low-energy isoscalar dipole response of heavy spherical nuclei is studied by using a semiclassical model, based on the solution of the linearized Vlasov kinetic equation for finite Fermi systems. In this translation-invariant model the excitations of the center of mass motion are exactly separated from the internal ones. The low-energy dipole strength function displays three resonance structures in the energy region up to 15 MeV. Calculations of the velocity fields associated with the resonance structures at the centroid energies show a vortex (toroidal) character of two overlying resonances, while the origin of the lowest isoscalar dipole resonance structure is related to the dipole single-particle excitations. Its centroid energy is close to the minimum energy of the dipole single-particle spectrum. The main toroidal resonance gives a qualitative description of the low-energy isoscalar dipole resonance observed in heavy spherical nuclei.

## 1. Introduction

The nuclear isoscalar dipole response reveals the low-energy resonance [1-4]. Theoretical studies of low-energy isoscalar dipole resonance had been carried out both within quantum approaches [5-11] and using semiclassical ones [12-16]. They showed that this resonance has a substantially vortex (toroidal) character. However, calculations of the isoscalar dipole strength function usually show several resonance structures in the energy region of low-energy resonanse (below 15 MeV). The main resonance is interpreted as the low-energy nuclear mode. It is of interest to study the nature of other dipole resonance structures in order to obtain additional information on the formation of low-energy isoscalar dipole resonances. For this, it is advisable to study the velocity fields associated with resonances.

In this paper, the collective isoscalar dipole excitations are considered within a translation-invariant kinetic model of small oscillations of finite Fermi systems [17]. In Section 2, we briefly recall the formalism of the kinetic model of collective dipole excitations in nuclei. The internal response function and the velocity field associated with the isoscalar dipole excitations are considered. In Section 3, the low-energy resonance structures of the isoscalar dipole response function are disccused, and the results of numerical calculations of the velocity fields associated with the low-energy resonance structures of the dipole strength function are shown.

## 2. Formalism

The translation-invariant kinetic model of small oscillations of finite Fermi systems based on the direct solution of the Vlasov equation for a Fermi system with moving surface is used to study the collective isoscalar dipole excitations of heavy nuclei [13, 15,16]. In this model, a nucleus is treated as a gas of interacting fermions confined to a spherical cavity with moving surface. Within our kinetic model, we can find the explicit expression for the fluctuation of the phase-space distribution function related to the collective isoscalar dipole excitations. By using this function, we can calculate the response function [15] as well as the local dynamical quantities, in particular, the velocity field [16].

Isoscalar dipole excitations in finite Fermi systems are an effect of the second order for the dipole moment (in the first order, they reduce to the center-of-mass motion). So, we consider the collective isoscalar dipole modes excited by a weak external field of the kind

$$V(\vec{r},t) = \beta \, \delta(t) \, Q^{(3)}(r) \, Y_{10}(\theta,\varphi), \tag{1}$$

where $Q^{(3)}(r) = r^3$ is the second-order dipole moment, $\delta(t)$ is the Dirac delta-function in time, and $\beta$ is a parameter that describes the external field strength. Within the kinetic model, assuming a simplified residual interaction of separable form,

$$v(\vec{r},\vec{r}') = \kappa_1 \sum_M r\, r' \, Y_{1M}(\theta,\phi) Y_{1M}^*(\theta',\phi') \tag{2}$$

the internal response function that is related with the collective isoscalar dipole excitations is given by

$$\tilde{R}_{intr}(s) = R_{33}(s) + S_{33}(s) - \tilde{R}_{c.m.}(s), \tag{3}$$

where $s = \omega R / v_F$ is a convenient dimensionless frequency ($v_F$ is the Fermi velocity). The function $R_{33}(s)$ is the collective fixed-surface response function, while $S_{33}(s)$ represents the moving-surface contribution. The function $\tilde{R}_{c.m.}(s)$ is the displacement of the nuclear center of mass induced by the external field (1) and is defined as

$$\tilde{R}_{c.m.}(s) = \frac{3}{8\pi} \frac{AR^6}{\epsilon_F s^2}, \tag{4}$$

where $\epsilon_F$ is the Fermi energy.

With the simple interaction (2) the function $R_{33}(s)$ can be evaluated explicitly as [15]

$$R_{33}(s) = R_{33}^0(s) + \kappa_1 \frac{[R_{13}^0(s)]^2}{1 - \kappa_1 R_{11}^0(s)}. \tag{5}$$

Here, the zero-order response functions $R_{jk}^0(s)$, $(j,k=1,3)$, are analogous to the single-particle response functions of the quantum theory and are given explicitly by [13]

$$R_{jk}^0(s) = \frac{9A}{16\pi} \frac{1}{\epsilon_F} \sum_{n=-\infty}^{+\infty} \sum_{N=\pm 1} \int_0^1 dx x^2 s_{nN}(x) \frac{Q_{nN}^j(x) Q_{nN}^k(x)}{s + i\varepsilon - s_{nN}(x)} \quad (j,k=1,3), \tag{6}$$

where the dimensionless single-particle angular momentum $x$ is $x = \sqrt{1-(l/p_F R)^2}$ and the dimensionless dipole single-particle eigenfrequencies $s_{nN}(x)$ are defined as

$$s_{nN}(x) = \frac{n\pi + N \arcsin(x)}{x}. \tag{7}$$

The quantity $\varepsilon$ is a vanishingly small parameter that determines the integration path at poles. The Fourier coefficients $Q_{nN}^k(x)$ are the classical limit of the quantum-mechanical radial matrix elements of the dipole operators and are given by

$$Q_{nN}^1(x) = (-)^n R \frac{1}{s_{nN}^2(x)}, \tag{8}$$

$$Q_{nN}^3(x) = 3R^2 Q_{nN}^1(x) \left(1 + \frac{4}{3} N \frac{\sqrt{1-x^2}}{s_{nN}(x)} - \frac{2}{s_{nN}^2(x)}\right). \tag{9}$$

The response functions (6) involve an infinite sum over n, however in practice it is sufficient to include only a few terms around n=0 in order to fulfill the energy-weighted sum rule with good accuracy.

The moving-surface contribution $S_{33}(s)$ to the internal response function (3) can be evaluated explicitly as

$$S_{33}(s) = -\frac{1}{1 - \kappa_1 R_{11}^0(s)} \frac{[\chi_3^0(s) - \chi_3^0(0)\kappa_1 R_{11}^0(s)]^2}{[-\chi_1(s)][1 - \kappa_1 R_{11}^0(s)] + \kappa_1 [\chi_1^0(s) - \chi_1^0(0)]^2}, \tag{10}$$

where the functions $\chi_k^0(s), (k=1,3)$, and $\chi_1(s)$ describe the dynamical surface effects and are defined as in Ref. [13]:

$$\chi_k^0(s) = \frac{9A}{8\pi} \sum_{n=-\infty}^{+\infty} \sum_{N=\pm 1} \int_0^1 dx x^2 s_{nN}(x) \frac{(-)^n Q_{nN}^{(k)}(x)}{s+i\varepsilon - s_{nN}(x)} \quad (k=1,3), \tag{11}$$

$$\chi_1(s) = -\frac{9A}{4\pi}\epsilon_F(s+i\varepsilon) \sum_{n=-\infty}^{+\infty} \sum_{N=\pm 1} \int_0^1 dx x^2 \frac{1}{s+i\varepsilon - s_{nN}(x)}. \tag{12}$$

The poles of the internal response function (3) determine the frequencies of collective isoscalar dipole modes. Neglecting residual interaction ($\kappa_1 = 0$) in Eqs. (5), (10), we obtain the internal response function (3) in the zeroth-order approximation. Due to a self-consistent coupling between the motion of nucleons and the moving surface, the collective isoscalar dipole excitations originate in our kinetic model already in the zeroth-order approximation.

To get the information about the origin of collective isoscalar dipole excitations, it is interesting to consider the velocity field associated with the dipole collective motion. This local dynamic quantity describes the spatial distribution of the average nucleon velocity during collective excitation and provides information on the nature of excitation. In our kinetic model, the time Fourier-transform of the velocity field is determined as

$$\vec{u}(\vec{r},\omega) = \frac{1}{m\rho_0} \int d\vec{p}\, \vec{p}\, \delta n(\vec{r},\vec{p},\omega), \tag{13}$$

where $\delta n(\vec{r},\vec{p},\omega)$ is the (Fourier transformed in time) fluctuations of the phase-space particle distribution induced by a weak external field (1) and $\rho_0$ is the nuclear equilibrium density. Choosing the Z axis in the direction of the external field, we will consider the velocity field in the meridian plane XZ that usually exploited in the RPA calculations [5, 20]. In this representation, the radius-vector of particle is $\vec{r} = (x, y=0, z)$ or $\vec{r} = (r, \theta, \varphi=0)$ in the spherical coordinates and the velocity field (13) can be written as

$$\vec{u}(r,\theta,\varphi=0,\omega) = u_x(r,\theta,\omega)\vec{e}_x + u_z(r,\theta,\omega)\vec{e}_z, \tag{14}$$

where $u_x(r,\theta,\omega)$ and $u_z(r,\theta,\omega)$ are the projections of the velocity field vector into the X and Z axes, respectively, and $\vec{e}_x, \vec{e}_z$ are unit vectors directed along these axes. The expressions for the functions $u_x(r,\theta,\omega)$ and $u_z(r,\theta,\omega)$ can be written as, see Ref. [16],

$$u_x(r,\theta,\omega) = \sqrt{\frac{3}{5}} Y_{21}(\theta,0) u_{12}(r,\omega), \tag{15}$$

$$u_z(r,\theta,\omega) = Y_{00}(\theta,0) u_{10}(r,\omega) - \sqrt{\frac{2}{5}} Y_{20}(\theta,0) u_{12}(r,\omega). \tag{16}$$

Here $Y_{lm}(\theta,0)$ are the spherical harmonics, while the radial functions $u_{10}(r,\omega)$ and $u_{12}(r,\omega)$ are defined as

$$u_{12}(r,\omega) = -i\sqrt{\frac{2}{3}}\pi\frac{1}{\rho_0}\frac{1}{r^2}\int d\epsilon \int dl\, l \sum_{N=-1}^{1} \left\{-i[\delta\tilde{n}_N^+(r,\epsilon,l,\omega) - \delta\tilde{n}_N^-(r,\epsilon,l,\omega)] + \right.$$
$$\left. + \frac{N}{2}\frac{l}{p(r,\epsilon,l)\,r}[\delta\tilde{n}_N^+(r,\epsilon,l,\omega) + \delta\tilde{n}_N^-(r,\epsilon,l,\omega)]\right\}, \tag{17}$$

$$u_{10}(r,\omega) = -i\sqrt{\frac{1}{3}}\pi\frac{1}{\rho_0}\frac{1}{r^2}\int d\epsilon \int dl\, l \sum_{N=-1}^{1} \left\{i[\delta\tilde{n}_N^+(r,\epsilon,l,\omega) - \delta\tilde{n}_N^-(r,\epsilon,l,\omega)] + \right.$$
$$\left. + N\frac{l}{p(r,\epsilon,l)\,r}[\delta\tilde{n}_N^+(r,\epsilon,l,\omega) + \delta\tilde{n}_N^-(r,\epsilon,l,\omega)]\right\}, \tag{18}$$

where $\epsilon$ is the particle energy, $l$ is the magnitude of its angular momentum, and $p(r,\epsilon,l) = p_F\sqrt{1-(l/p_F r)^2}$ is the magnitude of the particle radial momentum. The fluctuations of the phase-space particle distribution functions $\delta\tilde{n}_N^\pm(r,\epsilon,l,\omega)$ are the solutions of the linearized Vlasov kinetic equation for a finite system with moving surface. The explicit expressions of these functions are given in Ref. [16]. Our semiclassical projections (15) and (16) of the velocity field vector are similar to the quantum ones; see e.g., [18]. They have the same angular dependence as the quantum dipole velocity field, while the radial form factors are calculated using the RPA-type equations of motion.

## 3. Low-energy resonances

In Fig. 1 we display the low-energy part (up to 20 MeV) of the isoscalar dipole strength function ($E = \hbar\omega$)

$$S(E) = -\frac{1}{\pi}\mathrm{Im}\tilde{R}_{intr}(E). \tag{19}$$

The dashed curve is obtained from the internal response function (3) in the zero-order approximation ($\kappa_1 = 0$), while the solid curve shows the internal response function (3) taking into account the residual interaction between nucleons. It can be seen from Fig. 1 that the strength function has three resonance structures already in the zero-order approximation. The inclusion of the residual interaction leads to an insignificant shift of the resonance structures towards low energies. The strength parameter of the isoscalar dipole interaction (2), chosen in order to reproduce the experimental value of the giant monopole resonance energy in $^{208}Pb$ within our kinetic model [14], is $\kappa_1 = -7.5 \cdot 10^{-3}$ MeV/fm$^2$. The corresponding value of the incompressibility modulus equals $K_A = 160$ MeV [15]. The numerical calculations were carried out using standard values of nuclear parameters: $r_0 = 1.25$ fm, $\epsilon_F = 30.94$ MeV, and $m = 1.04$ MeV$(10^{-22}$s$)^2$/fm$^2$.

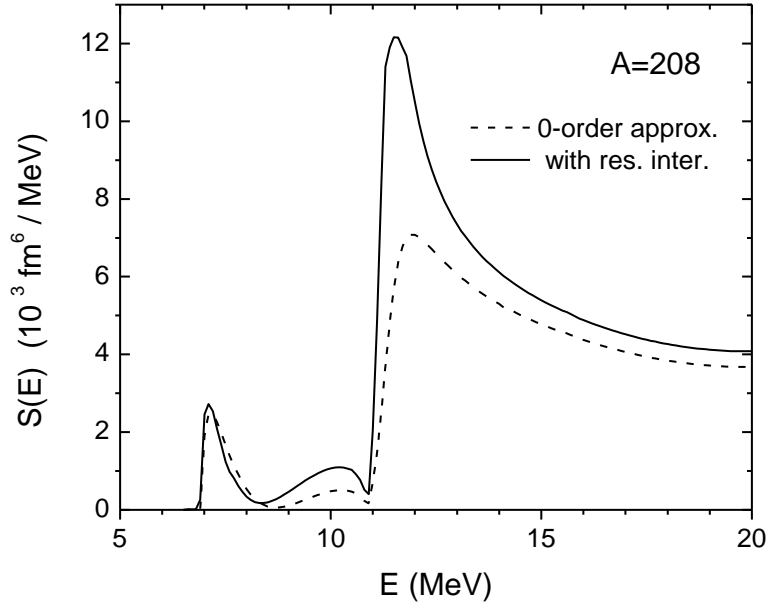

*Fig.1. Isoscalar dipole strength function in the low-energy region taking into account the residual interaction between nucleons (solid curve) and in the zero-order approximation (dashed curve). The system contains A=208 nucleons.*

We calculate the velocity field (14) associated with the resonance structures of the dipole strength function at the centroid energies.

In Fig. 2 the velocity fields for the lowest resonance structure are shown in the zero-order approximation at the centroid energy 7.2 MeV (a) and with regard for the residual interaction at the centroid energy 7.1 MeV (b). This resonance structure is associated with single-particle dipole excitations. Its centroid energy is close to the minimum energy of the dipole single-particle spectrum (7). Indeed, the lowest branch of single-particle dipole frequencies $s_{01}(x)$ has a gap that determines the minimum dimensionless frequency $s_{\min} = s_{01}(x=1) = 1$. Taking into account that $s = \omega R/v_F$, we obtain $E_{\min} = \hbar v_F/R$. Using the parameters of our model, we can find that $E_{\min}$ is approximately equal to the centroid energy of the lowest resonance structure.

On the other hand, the velocity fields associated with the overlying resonances of the strength function have the vortex (toroidal) character already in the zero-order approximation, see Figs. 3, 4.

In Fig. 3, the results of numerical calculations of the velocity fields associated with the second resonance of the strength function are shown in the zero-order approximation (a) and taking into account the residual interaction (b). It can be seen from Fig. 3 that the inclusion of the residual interaction leads to strengthening the vortex motion associated with this resonance. Fig. 4a shows the velocity field for the main toroidal resonance in the zero-order approximation at the centroid energy 12 MeV, while Fig. 4b displays this velocity field taking into account the residual interaction at the centroid energy 11.5 MeV. This resonance reproduces the nuclear low-energy resonance observed in heavy nuclei [16].

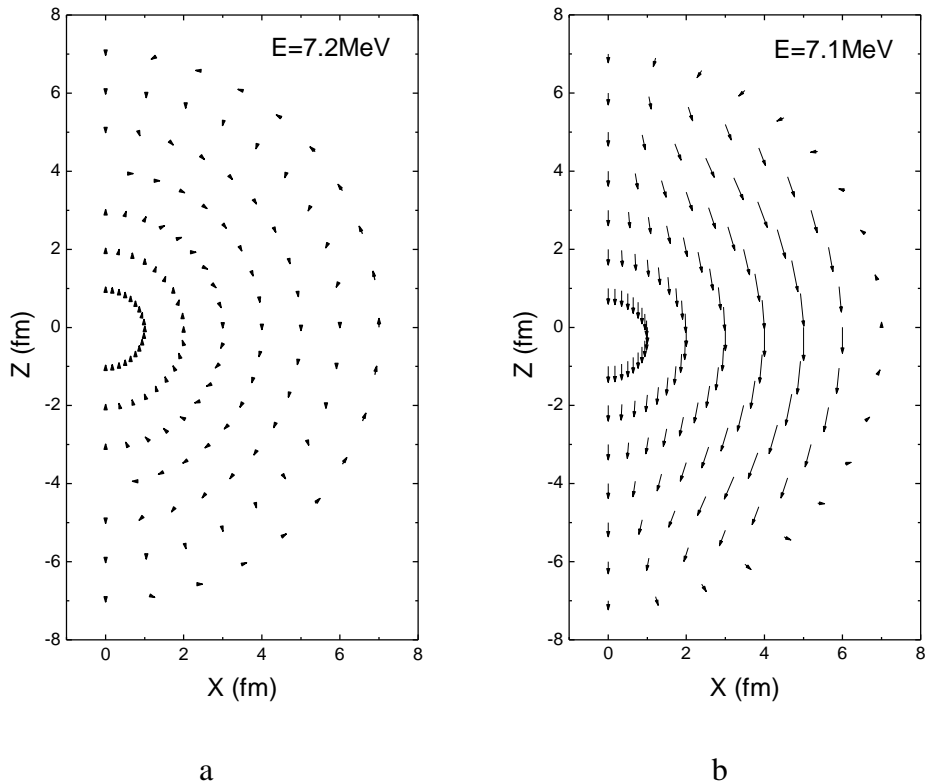

a  b

*Fig.2. The velocity fields in the XZ plane associated with the lowest resonance structure, see Fig. 1, in the zero-order approximation at the centroid energy of 7.2 MeV (a) and taking into account the residual interaction between the nucleons at the centroid energy of 7.1 MeV (b). The system contains A = 208 nucleons.*

## 4. Conclusions

The velocity fields associated with the low-energy resonance structures of the isoscalar dipole strength function have been studied within the kinetic model. In this model, taking into account the dynamical-surface degree of freedom, it is possible to obtain an exact treatment of the centre of mass motion. It is found that our semiclassical model predicts two toroidal resonances in the energy region below 15 MeV. The main toroidal resonance (the centroid energy 11.5 MeV), see Fig. 4b, reproduces the nuclear low-energy resonance. The neighboring resonance in the region of lower energies (the centroid energy 10.2 MeV), see Fig. 3b, also has a vortex (toroidal) character. The results of our semiclassical model are in qualitative agreement with the previous results of the relevant random-phase-approximation (RPA) calculations [5, 10, 11]. The quantitative comparison of our semiclassical model and the quantum approaches is rather difficult due to

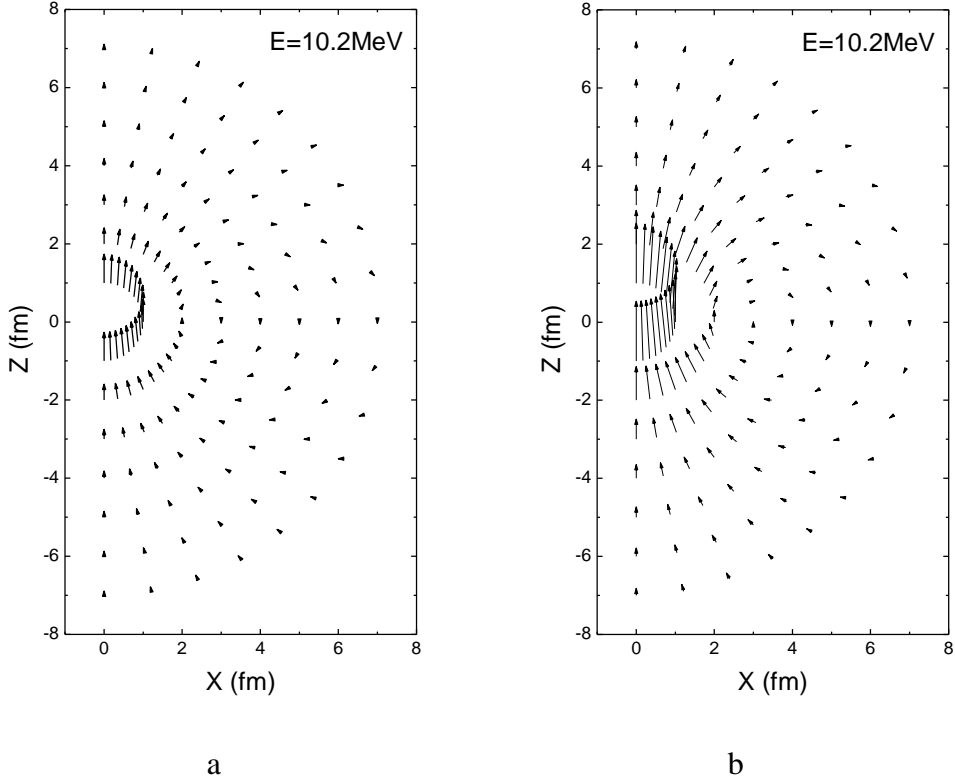

a                        b

*Fig.3. Velocity fields in the XZ-plane associated with the second resonance structure, see Fig.1, in the zero-order approximation at the centroid energy 10.2 MeV (a) and taking into account the residual interaction between nucleons at the centroid energy 10.2 MeV (b). The system contains A=208 nucleons.*

different nature of calculations. The lowest resonance structure (the centroid energy 7.1 MeV), see Fig. 2b, does not have collective excitation properties. Its centroid energy is close to the minimum energy of the dipole single-particle spectrum ($E_{\min} = \hbar\, v_F / R$). The nature of the velocity field associated with this resonance structure is determined by the direction of the external force (1).

Our semiclassical approach makes it possible to obtain additional information on the nature of collective isoscalar dipole excitations in heavy nuclei. In particular, it would be interesting to study the nature of the momentum flux associated with collective isoscalar dipole excitations. This study could clearly show the effect of dynamic deformation of the Fermi surface on the formation of nuclear low-energy resonance.

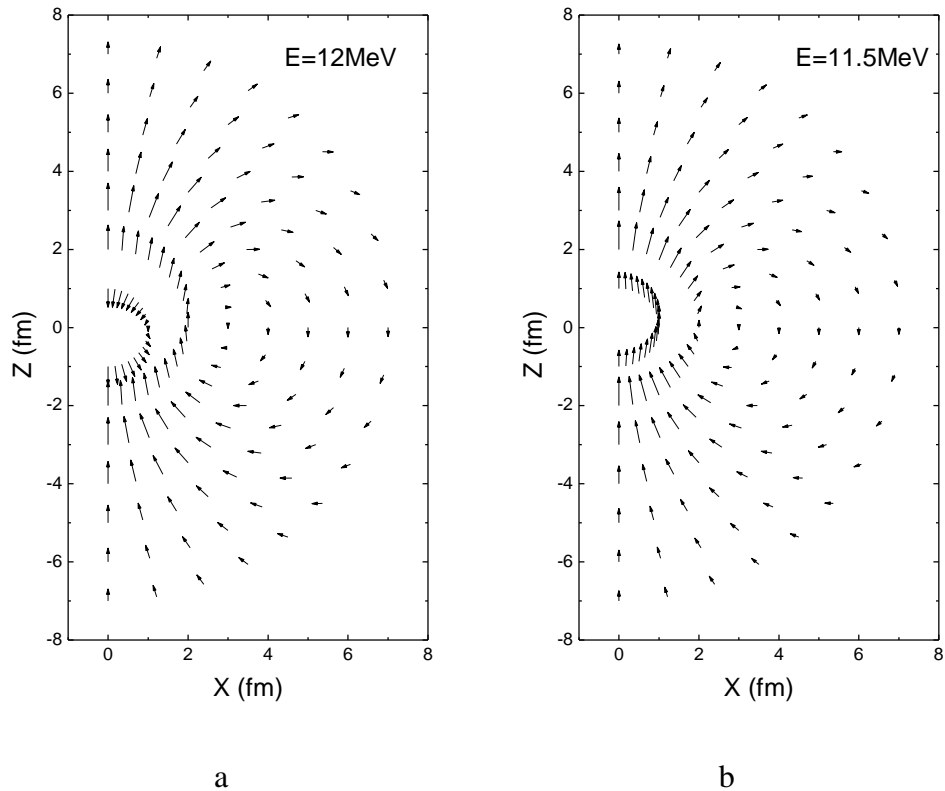

a　　　　　　　　　　　　　　　b

*Fig.4. Velocity fields in the XZ-plane associated the third resonance structure, see Fig.1, in the zero-order approximation at the centroid energy 12 MeV (a) and taking into account the residual interaction between nucleons at the centroid energy 11.5 MeV (b). The system contains A=208 nucleons.*